\documentclass[11pt,letterpaper]{article}
\usepackage[T1]{fontenc}
\usepackage{lmodern}
\usepackage[margin=1.08in]{geometry}
\usepackage{amsmath,amssymb,amsthm}
\usepackage{microtype}
\usepackage{enumitem}
\usepackage{xcolor}
\usepackage{tikz}
\usetikzlibrary{arrows.meta}
\definecolor{linkblue}{RGB}{30,58,91}
\usepackage[colorlinks=true,allcolors=linkblue,bookmarksnumbered=true,
  pdfauthor={Cheng Du and Zongqi Wan},
  pdftitle={Random Gaps for Optimal Average-Case Shellsort}]{hyperref}
\newtheorem{theorem}{Theorem}[section]
\newtheorem{corollary}[theorem]{Corollary}
\newtheorem{lemma}[theorem]{Lemma}
\newcommand{\E}{\mathbb E}
\newcommand{\I}{\mathcal I}
\newcommand{\Sn}{\mathfrak S_n}
\newcommand{\pos}[1]{(#1)_+}

\newcommand{\heading}[1]{\par\smallskip\noindent\textbf{#1}\enspace}
\setlist[enumerate]{topsep=4pt,itemsep=3pt,parsep=0pt,leftmargin=*}
\title{Shellsort with $O(n\log n)$ Average-Case Complexity}
\author{Cheng Du\thanks{Great Bay University, Email: cdu@gbu.edu.cn} \and Zongqi Wan\thanks{Great Bay University, Email: zqwan@gbu.edu.cn}}
\date{}

\begin{document}
\maketitle
\begin{abstract}
We show that Shellsort admits a fixed sequence of $O(\log n)$ gaps that sorts a uniform random permutation in $O(n\log n)$ expected comparisons. In fact, we give a distribution on decreasing gap sequences with this expected bound for every fixed input. We also give a randomized polynomial-time algorithm that, with high probability, constructs a sequence attaining the average-case bound.
\end{abstract}

\section{Introduction}
Shellsort is a classic sorting method which generalizes insertion sort~\cite{shell,knuth}. In ordinary insertion sort, a key moves left through the sorted prefix one position at a time. Shell's idea was to start with a larger step, called a \emph{gap}. A pass with gap $h$ applies insertion sort separately to each subsequence of positions $r,r+h,r+2h,\ldots$, so a single exchange within a subsequence moves a key $h$ positions in the array. For example, a pass with gap $2$ sorts the keys in the odd positions and those in the even positions separately. Shellsort uses successively smaller gaps, allowing early passes to move keys over long distances before a final pass with gap $1$ completes the sort.

How should the gaps be chosen to minimize the average number of comparisons on a random permutation? For two passes even the optimal first gap gives order $n^{5/3}$ comparisons. At the other extreme, Pratt's sequence gives $O(n\log^2 n)$ comparisons on every input~\cite{knuth,pratt}. Sedgewick~\cite{sedsurvey} asked whether some sequence attains $O(n\log n)$ on average.

Jiang, Li, and Vit\'anyi~\cite{jlv} identified the only possible order of the number of passes. For $p$ passes their lower bound on uniform inputs is $\Omega(pn^{1+1/p})$ comparisons in the range $1\le p\le n/2$. Too few passes make the factor $n^{1/p}$ expensive. Too many make the passes themselves expensive. The two costs allow an $O(n\log n)$ average only at $p=\Theta(\log n)$. The lower bound leaves open whether there exists a fixed sequence of $\Theta(\log n)$ gaps for which Shellsort uses $O(n\log n)$ comparisons on average over uniformly random permutations.

For a permutation $A\in\Sn$ and gap sequence $H$, let $C(H,A)$ be the total number of key comparisons, including the comparisons that stop an insertion.
\begin{theorem}\label{thm:main}
For each $n\ge2$ there is an explicitly specified distribution on strictly decreasing sequences of positive integer gaps ending in $1$, of length $O(\log n)$, such that
\[
   \max_{A\in\Sn}\ \E_H C(H,A)=O(n\log n).
\]
Every realization sorts every input.
\end{theorem}

By averaging over uniformly random inputs, Theorem 1.1 implies the existence of a fixed gap sequence with optimal average-case comparison complexity.

\begin{corollary}\label{cor:fixed}
For every $n\ge2$ there is a fixed sequence $H_n$ of $O(\log n)$ gaps such that
$\E_{A\sim \mathrm{Unif}(\Sn)} C(H_n,A)=\Theta(n\log n)$.
\end{corollary}

Corollary~\ref{cor:fixed} give the existence of the gap sequence with $O(n\log n)$ average-case complexity. For a fixed $H$, write $\overline C(H)\triangleq \E_{A\sim\mathrm{Unif}(\Sn)} C(H,A)$. By Theorem~\ref{thm:main} and Markov's inequality, a single draw of $H$ satisfies $\overline C(H)\le c_0 n\log n$ with probability at least $1/2$ for an absolute $c_0$.
With extra sampling and scoring technique, we can efficiently obtain such a desired gap sequence with arbitrarily high probability. The resulting sequence can then be reused on any input of length $n$.
\begin{theorem}[Sequence construction]\label{thm:construct}
Given $n\ge2$ and $0<\delta<1$, there exist a randomized algorithm outputs a strictly decreasing sequence $H$ of $O(\log n)$ gaps ending in $1$, in time polynomial in $n$ and $\log(1/\delta)$. For an absolute constant $c$,
\[
    \Pr\bigl[\,\overline C(H)\le c\,n\log n\,\bigr]\ge 1-\delta .
\]
The probability is over the construction of $H$.
\end{theorem}

\subsection{Random Gap Schedule}\label{sec:schedule}
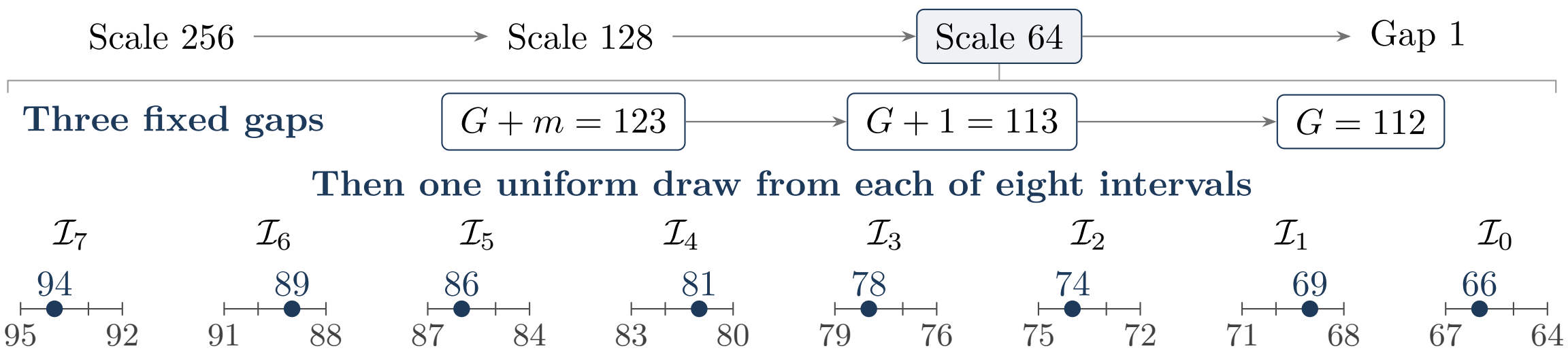
\begin{figure}[htb]
\centering
\begin{tikzpicture}[
  x=.064\linewidth, y=1cm, font=\small,
  flow/.style={-{Stealth[length=4pt]}, draw=black!55, thin},
  stage/.style={inner xsep=5pt, inner ysep=4pt},
  group/.style={font=\small\bfseries, text=linkblue},
  endpoint/.style={font=\footnotesize, text=black!75, inner sep=1pt}
]
  \node[stage] (initial) at (1.45,2.75) {Scale $256$};
  \node[stage] (half) at (5.40,2.75) {Scale $128$};
  \node[stage,draw=linkblue,fill=linkblue!7,rounded corners=2pt]
    (current) at (9.35,2.75) {Scale $64$};
  \node[stage] (finish) at (13.30,2.75) {Gap $1$};
  \foreach \prevstage/\nextstage in
    {initial/half,half/current,current/finish}
    \draw[flow] (\prevstage.east) -- (\nextstage.west);

  \draw[black!40] (current.south) -- (9.35,2.32)
    (0,2.20) -- (0,2.32) -- (14.60,2.32) -- (14.60,2.20);
  \node[group,anchor=west] at (0,1.92) {Three fixed gaps};
  \node[stage,draw=linkblue,rounded corners=2pt] (fixedone) at (5.24,1.92) {$G+m=123$};
  \node[stage,draw=linkblue,rounded corners=2pt] (fixedtwo) at (9.00,1.92) {$G+1=113$};
  \node[stage,draw=linkblue,rounded corners=2pt] (fixedthree) at (12.76,1.92) {$G=112$};
  \draw[flow] (fixedone.east) -- (fixedtwo.west);
  \draw[flow] (fixedtwo.east) -- (fixedthree.west);
  \node[group] at (7.30,1.29) {Then one uniform draw from each of eight intervals};

  \foreach \startx/\slot/\highgap/\lowgap/\offset/\chosen in
    {.12/7/95/92/1/94,2.04/6/91/88/2/89,3.96/5/87/84/1/86,5.88/4/83/80/2/81,
     7.80/3/79/76/1/78,9.72/2/75/72/1/74,11.64/1/71/68/2/69,13.56/0/67/64/1/66}{
    \begin{scope}[shift={(\startx,0)}]
      \node at (.48,.82) {$\I_{\slot}$};
      \draw[black!70] (0,.10) -- (.96,.10);
      \foreach \tick in {0,...,3}
        \draw[black!70] (.32*\tick,.035) -- (.32*\tick,.165);
      \node[endpoint,below] at (0,-.015) {$\highgap$};
      \node[endpoint,below] at (.96,-.015) {$\lowgap$};
      \fill[linkblue] (.32*\offset,.10) circle[radius=2.3pt];
      \node[above,text=linkblue,inner sep=1.8pt] at (.32*\offset,.16) {$\chosen$};
    \end{scope}
  }
\end{tikzpicture}
\caption{The schedule for $n=1024$. At the last scale, $s=64$, $G=112$, and $m=11$.
The three fixed gaps are followed by eight draws from intervals of width $w=4$
partitioning $[64,96)$. Each dot marks the gap chosen for one pass, in execution order from left to right.}
\label{fig:gap-sequence}
\end{figure}

Figure~\ref{fig:gap-sequence} shows one scale of the construction. At scale $s$, we first use three fixed gaps near $7s/4$, then choose one random gap from each of eight consecutive intervals below $3s/2$. The fixed gaps put sufficiently distant pairs in order. The random gaps reduce the disorder that remains before we halve $s$.

Set $s_*\triangleq 64$. If $n<256$, use only gap $1$. Otherwise let $s_0$ be the largest power of two at most $n/4$. For $s=s_0,s_0/2,\ldots,s_*$, perform the following passes in order.
\begin{enumerate}
\item \textbf{Fixed gaps.} Set $G\triangleq 7s/4$ and $m\triangleq \lceil\sqrt G\rceil$. Use the gaps $G+m,\,G+1,\,G$.
\item \textbf{Random gaps.} Partition the integers in $[s,3s/2)$ into eight intervals of width $w\triangleq s/16$,
\[
         \I_t\triangleq \{s+tw,\ s+tw+1,\ldots,s+(t+1)w-1\},\qquad 0\le t<8.
\]
Visit the intervals from $\I_7$ down to $\I_0$. In each interval draw one integer uniformly and use it as the gap for one pass. Make all draws independently.
\end{enumerate}
After the last scale, use gap $1$. Since $s$ is a power of two at least $64$, $w$ is an integer and $1<m<s/4$. All gaps at this scale lie in $[s,2s)$. Visiting disjoint intervals in descending order keeps the sequence strictly decreasing for every draw, and the next scale starts below $s$. For $n\ge256$, the eleven passes per scale and the final pass give exactly $11(1+\log_2(s_0/64))+1$ passes.

\subsection{Technical Overview}\label{sec:overview}
We fix an input permutation and analyze the dyadic scales of the schedule in Figure~\ref{fig:gap-sequence}. At each scale, we show that the passes cost $O(n)$ expected comparisons and reduce the remaining disorder enough to proceed at half the scale.

\heading{Potential and comparison cost.}
Let $A_i$ be the rank of the key at position $i$, so its sorted position is $A_i$. We measure disorder by $Q(A)\triangleq \sum_i(A_i-i)^2$. We will show that, an $h$-pass never increases this potential and uses at most
$
                         n+\frac{2}{h}\sqrt{nQ(A)}
$
comparisons (Lemma~\ref{lem:cost}). We maintain $\E Q(A)\le Dns^2$ on entry to scale $s$, with a common constant $D$. Since every gap at this scale is at least $s$, the bound gives $O(n)$ expected comparisons per pass, which sums to desired $O(n \log n)$ average-case complexity.
So the only remaining question is how to keep this invariant during each scale. Note that, scales decrease dyadically. Thus, we need the potential to be quartered when a scale ends.

\heading{Inversion weight and distance.}
For each position pair $(i,j)$ with $i<j$, assign weight $v_{ij}\triangleq \max\{A_i-A_j,0\}$ to it. These weights sum to $\frac{Q(A)}{2}$. Writing $W_h(A)$ for the weight of pairs in the same $h$-chain, we prove in Lemma~\ref{lem:cost}
\[
                         Q(A)-Q(P_h\circ A)=2hW_h(A).
\]
Since $h=\Theta(s)$ at scale $s$, capturing an $\Omega(1/s)$ fraction of the total inversion weight in expectation gives a constant-factor expected decrease in $Q(A)$. To prove such a guarantee, we first locate where this weight can remain. Each pass preserves sortedness at every previously used gap (Lemma~\ref{lem:preserve}). Thus, after the three fixed passes, the array is simultaneously sorted at all three gaps, and later passes preserve these properties. By transitivity, pairs whose distance is a nonnegative integer combination of these gaps are also ordered. The chosen gaps represent every distance beyond a constant multiple of $s^{3/2}$ in this way, so all inversions at these large distances are eliminated and stay absent throughout the scale (Lemma~\ref{lem:far}).

We next bound the weight that can remain at short distances. In the proof of Lemma~\ref{lem:drift}, we show that inversions at distance below a sufficiently large constant multiple of $s$ have total weight $O(s\sqrt{nQ(A)})$. When $Q(A)$ exceeds a sufficiently large multiple of $ns^2$, this is only a small fraction of the total weight $Q(A)/2$. Since inversions with large distance are absent by Lemma~\ref{lem:far}, a constant fraction of the weight must then lie at intermediate distances. Thus, unless the potential is already $O(ns^2)$, it suffices to analyze how the random passes capture this intermediate weight.

\heading{Potential Progress from a random gap.}
For an intermediate distance $d$, we count representations $d=kh+(k+1)g$, with $k\ge1$ and $h,g$ in the same candidate interval. Each representation splits the pair into two shorter pairs, one in an $h$-chain and one in a $g$-chain. Their weights sum to at least the original weight. Each distance has $\Omega(s)$ representations (Lemma~\ref{lem:representations}), while each position pair with a fixed gap label is used by $O(s)$ paths across all endpoint pairs. The consecutive multipliers control this overlap, since either segment determines $k$ and leaves only the other gap to choose.

Let $M_s$ denote the total weight of intermediate inversions, and let $\I$ be the candidate interval for the current pass. Averaging the weight inequality over representations and using the bound on reuse gives $\sum_{q\in\I}W_q(A)=\Omega(M_s)$.
Since $\I$ contains $\Theta(s)$ candidates, a uniform draw captures expected weight
\[
    \E_h W_h(A)=\frac{1}{|\I|}\sum_{q\in\I}W_q(A)
              =\Omega\!\left(\frac{M_s}{s}\right).
\]
To convert this captured weight into a potential decrease, apply the identity from Lemma~\ref{lem:cost}. Every candidate gap satisfies $h\ge s$, so
\[
    \E_h\!\left[Q(A)-Q(P_h\circ A)\right]
       =\E_h[2hW_h(A)]
       \ge 2s\,\E_h W_h(A)=\Omega(M_s).
\]
Thus the multiplier $h$ in the potential-drop identity cancels the $\frac{1}{s}$ loss from choosing only one candidate gap. Lemma~\ref{lem:routing} gives the quantitative form of this decrease. Combining it with the short-distance bound $M_s\ge\frac{Q(A)}{2}-O(s\sqrt{nQ(A)})$ yields the following contraction (Lemma~\ref{lem:drift}):
\[
                    \E_h Q(P_h\circ A)\le0.77\,Q(A)+O(ns^2).
\]

After eight random passes, the coefficient of the incoming potential is at most $0.77^8<\frac{1}{8}$, with an additive $O(ns^2)$ term. Taking $D$ sufficiently large restores the bound $\E Q(A_{\rm out})\le Dn(\frac{s}{2})^2$ for the array leaving the scale. There are $O(\log n)$ scales, and the final gap-$1$ pass at a constant cutoff costs $O(n)$ in expectation. This proves Theorem~\ref{thm:main}.

\heading{Constructing a fixed gap sequence.}
Averaging over inputs shows the existence of a fixed sequence of optimal average cost. To construct such a sequence efficiently, we estimate each candidate gap's mean potential on random test permutations and choose an empirical minimizer. Polynomially many samples give enough accuracy to preserve the contraction needed.

\subsection{Related Work}\label{sec:related}
\heading{Average-Case Bounds for Fixed Gap Sequences.}
Espelid~\cite{espelid} analyzed the original power-of-two schedule for power-of-two input lengths, a logarithmic-pass example with average exchange count of order $n^{3/2}$. This illustrates why the right number of gaps alone is insufficient. Knuth's analysis of two-pass Shellsort gives optimal average cost $\Theta(n^{5/3})$ with a first gap of order $n^{1/3}$~\cite[Sec.~5.2.1]{knuth}. For three passes Yao~\cite{yao} derived a combinatorial description of the exchanges in the final insertion sort. Janson and Knuth~\cite{jk} developed this analysis into an $O(n^{23/15})$ average bound for a particular three-increment family. Vit\'anyi~\cite{vit} obtained a matching lower bound for that family. These results track what a few prescribed passes leave to the final pass.

For a general sequence, Jiang, Li, and Vit\'anyi~\cite{jlv} encode the input by the numbers of larger predecessors crossed by each key at each pass. Counting the possible encodings gives their pass-dependent lower bound. Vit\'anyi~\cite{vit} subsequently derived a sequence-sensitive lower bound $\Omega(n\sum_{k=1}^p h_{k-1}/h_k)$ on average exchanges, with $h_0\triangleq n$.

\heading{Arithmetic constructions and worst-case bounds.}
Pratt~\cite{pratt} used all gaps of the form $2^a3^b<n$ to give a worst-case $O(n\log^2 n)$ bound. Sedgewick~\cite{sed86} used the Frobenius problem to obtain an $O(n^{4/3})$ worst-case bound with logarithmically many gaps, and Weiss and Sedgewick~\cite{ws} proved this bound tight for that sequence. Incerpi and Sedgewick~\cite{is} obtained $n^{1+o(1)}$ worst-case bounds with logarithmically many passes. Selmer~\cite{selmer} studied the connection to the Frobenius problem directly.

Poonen~\cite{poonen} gave worst-case lower bounds depending on the number of gaps. Plaxton, Poonen, and Suel~\cite{pps} and Plaxton and Suel~\cite{ps} obtained the general worst-case lower bound $\Omega(n(\log n/\log\log n)^2)$. Cypher~\cite{cypher} studied the size of sorting networks with decreasing Shellsort increments. More recently, Zang~\cite{zang} proved worst-case lower bounds for gaps close to rational geometric sequences, and Zhao and Ni~\cite{zhaoni} studied how previously used gaps constrain the exchanges in a subsequent pass.

\heading{Empirical gap selection and randomized Shellsort.}
Weiss~\cite{weiss} measured the growth of the average comparison count for several sequences. Ciura~\cite{ciura} used sequential analysis to search for increments with small average costs at finite input sizes. Hu~\cite{hu} found empirical fits consistent with $O(n\log n)$ comparisons for a particular gap sequence. Liu~\cite{liu} used an execution-guided search to design a sparse sequence and analyzed its worst-case cost. Sanders and Fleischer~\cite{sf} and McGeoch et al.~\cite{mcgeoch} tested classical Shellsort with offsets formed from products of random numbers and studied the resulting comparison counts.

Goodrich's randomized Shellsort~\cite{goodrich} achieves $O(n\log n)$ work and sorts each fixed input with very high probability by comparing elements paired by random matchings between regions. A Las Vegas conversion also gives a high-probability time bound. Earlier variants, including those analyzed by Brejov\'a~\cite{brejova}, replace chain insertion sorts by bubble or shaker passes.

\section{Analysis of the Random Gap Sequence}\label{sec:analysis}
We prove Theorem~\ref{thm:main} and Corollary~\ref{cor:fixed} in this section.
Throughout our analysis, we define several quantities to capture the disorderness of the permutation being processed.
Let $P_h \circ A$ denote the result of sorting each residue class modulo $h$. For $i<j$ set $v_{ij}\triangleq \pos{A_i-A_j}$. With $\delta_i\triangleq A_i-i$, define
\[
 F(A)\triangleq \sum_i|\delta_i|,\qquad Q(A)\triangleq \sum_i\delta_i^2,\qquad
 W_h(A)\triangleq \sum_{\substack{i<j\\h\mid(j-i)}}v_{ij}.
\]
The condition $h\mid(j-i)$ selects pairs of positions in the same $h$-chain. Thus, $W_h(A)$ is the weighted inversions within the chains with gap $h$.
Write $C_h(A)$ for the comparison count of this pass on $A$.

\subsection{Properties of the Potential and Sortedness}\label{sec:pass}
\begin{lemma}\label{lem:cost}
For every permutation and $1\le h\le n$, the following hold.
\begin{itemize}[leftmargin=2.5em,topsep=4pt,itemsep=3pt,parsep=0pt]
\item[(i)] $Q(A)-Q(P_h \circ A)=2hW_h(A)$.
\item[(ii)] $\sum_{i<j}v_{ij}=Q(A)/2$.
\item[(iii)] $C_h(A)\le n+2\sqrt{nQ(A)}/h$.
\end{itemize}
\end{lemma}
\begin{proof}
View the pass as insertion sort by adjacent exchanges inside each $h$-chain. Each exchange swaps an inverted pair of keys $x>y$ at positions $p$ and $p+h$, placing $y$ at $p$ and $x$ at $p+h$. All other terms of the potential remain unchanged, so the decrease is
\[
 (x-p)^2+(y-p-h)^2-(y-p)^2-(x-p-h)^2=2h(x-y).
\]
Within a chain, the relative order of two keys can change only when those two keys are exchanged. Every initially inverted pair must change its relative order, since the chain is sorted at the end. Once exchanged, the pair is in increasing order and cannot be exchanged again, because the algorithm only swaps inverted pairs. For the same reason, pairs initially in increasing order are never exchanged. Thus an initial inversion at positions $i<j$ in the same chain contributes exactly $2h(A_i-A_j)=2hv_{ij}$ to the total decrease. Summing over all exchanges gives
\[
 Q(A)-Q(P_h\circ A)
    =2h\sum_{\substack{i<j\\h\mid(j-i)}}v_{ij}
    =2hW_h(A),
\]
which proves \textup{(i)}.

For \textup{(ii)}, set $h=1$ in \textup{(i)}. This sorts the entire permutation, so $Q(P_1 \circ A)=0$ and the identity follows.

For \textup{(iii)}, let $M_h$ be the number of initially inverted pairs in all chains. Each successful comparison crosses one such pair, and each insertion uses at most one additional key comparison. Thus $C_h(A)\le n+M_h$. If $i<j$ is an inverted pair at distance $d$, then $\delta_i-\delta_j=d+A_i-A_j>d$. Charge the pair to $i$ if $\delta_i\ge d/2$, and to $j$ otherwise. An index with positive displacement receives charges only from positions to its right, and an index with negative displacement only from its left. In either direction the eligible distances are multiples of $h$ at most $2|\delta_i|$. Hence $M_h\le 2F(A)/h$, and Cauchy--Schwarz gives $F(A)\le\sqrt{nQ(A)}$.
\end{proof}

Call $A$ \emph{$k$-sorted} if $A_i\le A_{i+k}$ whenever $i+k\le n$. We recall the classical preservation property~\cite{knuth,pratt} and include the boundary argument for chains of unequal lengths.
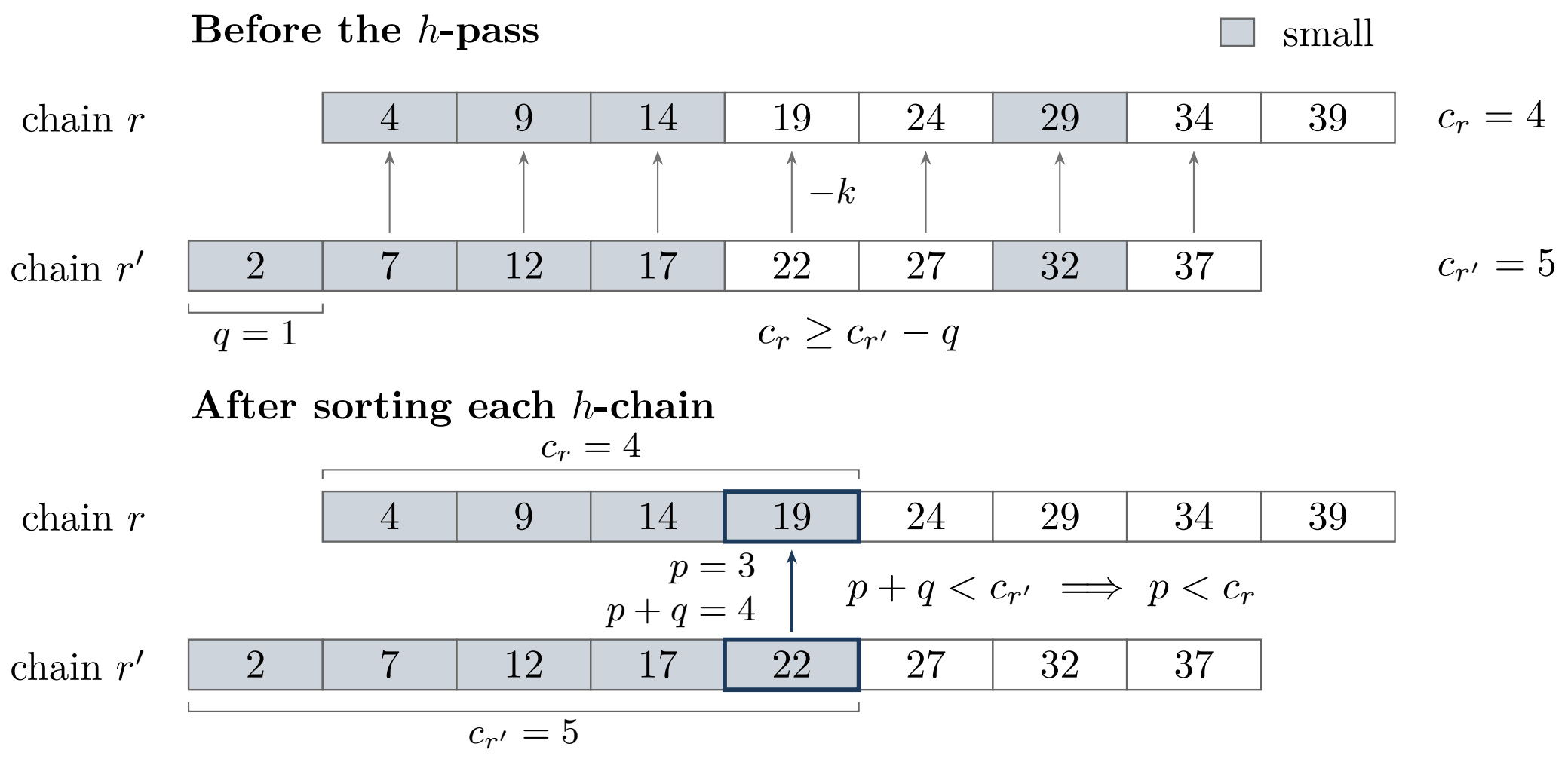
\begin{figure}[t]
\centering
\begin{tikzpicture}[
  x=.073\linewidth, y=1cm, font=\small,
  cell/.style={draw=black!60, fill=white, line width=.45pt},
  smallcell/.style={cell, fill=linkblue!22},
  predecessor/.style={-{Stealth[length=3.5pt]}, draw=black!55, thin},
  chainlabel/.style={anchor=east, inner sep=2pt},
  count/.style={font=\footnotesize, inner sep=2pt}
]
  \node[anchor=west,font=\small\bfseries] at (-.1,5.2) {Before the $h$-pass};
  \draw[smallcell] (7.7,5.08) rectangle (7.95,5.32);
  \node[anchor=west] at (8.05,5.2) {small};

  \foreach \slot/\position/\appearance in
    {0/4/smallcell,1/9/smallcell,2/14/smallcell,3/19/cell,
     4/24/cell,5/29/smallcell,6/34/cell,7/39/cell}{
    \draw[\appearance] (\slot+1,4.23) rectangle (\slot+2,4.67);
    \node at (\slot+1.5,4.45) {$\position$};
  }
  \foreach \slot/\position/\appearance in
    {0/2/smallcell,1/7/smallcell,2/12/smallcell,3/17/smallcell,
     4/22/cell,5/27/cell,6/32/smallcell,7/37/cell}{
    \draw[\appearance] (\slot,2.93) rectangle (\slot+1,3.37);
    \node at (\slot+.5,3.15) {$\position$};
  }
  \node[chainlabel] at (-.25,4.45) {chain $r$};
  \node[chainlabel] at (-.25,3.15) {chain $r'$};
  \node[anchor=west] at (9.2,4.45) {$c_r=4$};
  \node[anchor=west] at (9.2,3.15) {$c_{r'}=5$};
  \foreach \column in {1,...,7}
    \draw[predecessor] (\column+.5,3.44) -- (\column+.5,4.16);
  \node[count,fill=white,anchor=west] at (4.55,3.8) {$-k$};
  \draw[black!60] (0,2.82) -- (0,2.74) -- (1,2.74) -- (1,2.82);
  \node[count,below] at (.5,2.74) {$q=1$};
  \node at (5,2.55) {$c_r\ge c_{r'}-q$};

  \node[anchor=west,font=\small\bfseries] at (-.1,1.9) {After sorting each $h$-chain};
  \foreach \slot/\position/\appearance in
    {0/4/smallcell,1/9/smallcell,2/14/smallcell,3/19/smallcell,
     4/24/cell,5/29/cell,6/34/cell,7/39/cell}{
    \draw[\appearance] (\slot+1,.73) rectangle (\slot+2,1.17);
    \node at (\slot+1.5,.95) {$\position$};
  }
  \foreach \slot/\position/\appearance in
    {0/2/smallcell,1/7/smallcell,2/12/smallcell,3/17/smallcell,
     4/22/smallcell,5/27/cell,6/32/cell,7/37/cell}{
    \draw[\appearance] (\slot,-.57) rectangle (\slot+1,-.13);
    \node at (\slot+.5,-.35) {$\position$};
  }
  \node[chainlabel] at (-.25,.95) {chain $r$};
  \node[chainlabel] at (-.25,-.35) {chain $r'$};
  \draw[black!60] (1,1.28) -- (1,1.36) -- (5,1.36) -- (5,1.28);
  \node[count,above] at (3,1.36) {$c_r=4$};
  \draw[black!60] (0,-.68) -- (0,-.76) -- (5,-.76) -- (5,-.68);
  \node[count,below] at (2.5,-.76) {$c_{r'}=5$};
  \draw[linkblue,line width=1.1pt] (4,.73) rectangle (5,1.17);
  \draw[linkblue,line width=1.1pt] (4,-.57) rectangle (5,-.13);
  \draw[predecessor,draw=linkblue,line width=.8pt] (4.5,-.06) -- (4.5,.66);
  \node[count,anchor=east] at (4.3,.49) {$p=3$};
  \node[count,anchor=east] at (4.3,.11) {$p+q=4$};
  \node[anchor=west] at (4.8,.3) {$p+q<c_{r'}\ \Longrightarrow\ p<c_r$};
\end{tikzpicture}
\caption{The proof of Lemma~\ref{lem:preserve}, illustrated with $n=40$, $h=5$, $k=3$, $r=4$, $r'=2$, and $q=1$. Cells are labeled by array position. Shading marks entries of true rank at most $T$. Arrows point to predecessors $k$ positions earlier. The first $q$ positions of chain $r'$ are unmatched. Sorting preserves the counts and makes the small entries prefixes. The outlined pair shows how a small entry at $i+k$ forces a small entry at $i$ after the pass.}
\label{fig:sortedness-preservation}
\end{figure}

\begin{lemma}\label{lem:preserve}
If $A$ is $k$-sorted, then so is $P_h \circ A$, for every $h$.
\end{lemma}
\begin{proof}
Let $B\triangleq P_h\circ A$. Fix a rank threshold $T$ and call an entry \emph{small} if its true rank is at most $T$. Since $A$ is $k$-sorted, $A_i\le A_{i+k}$ whenever $i+k\le n$. Thus, before the pass, a small entry at $i+k$ implies a small entry at $i$. We will show that this implication also holds after the pass, for every threshold $T$. Figure~\ref{fig:sortedness-preservation} illustrates the chain correspondence and the effect of the pass.

The $h$-chain starting at $r$, $1\le r\le h$, consists of the valid positions $r+th$, $t=0,1,\ldots$. Let $c_r$ be its number of small entries. To compare two chains whose positions differ by $k$, write
\[
                 r+k=r'+qh,\qquad 1\le r'\le h,\quad q\ge0.
\]
Adding $k$ to position $r+ph$ on chain $r$ gives position $r'+(p+q)h$ on chain $r'$, whenever this position is within the array. Conversely, every valid position $j\triangleq r'+th$ with $t\ge q$ has a predecessor
\[
                 j-k=r+(t-q)h
\]
on chain $r$. This predecessor is a valid array position because it is at least $r\ge1$ and smaller than $j\le n$. If the entry at $j$ is small before the pass, then its predecessor is small as well, by $k$-sortedness. Distinct positions $j$ give distinct predecessors. Only the first $q$ positions of chain $r'$ can lack such a predecessor, so at least $c_{r'}-q$ of its small entries have distinct small predecessors on chain $r$. Therefore
\[
                              c_r\ge c_{r'}-q.
\]
This counting argument applies even when the two chains have different lengths.

Sorting a chain preserves its number of small entries and moves them to its first $c_r$ positions. Thus, after the pass, position $r+ph$ contains a small entry exactly when $p<c_r$. Now take any $i\triangleq r+ph$ with $i+k\le n$, so that $i+k=r'+(p+q)h$. If the entry at $i+k$ is small in $B$, then
\[
                 p+q<c_{r'}\quad\Longrightarrow\quad
                 p<c_{r'}-q\le c_r.
\]
Hence the entry at $i$ is small in $B$ as well.

Finally, choose $T\triangleq B_{i+k}$, the true rank of the key at position $i+k$ after the pass. The implication just proved gives $B_i\le B_{i+k}$. This holds for every $i$ with $i+k\le n$, so $B$ is $k$-sorted.
\end{proof}

\begin{lemma}\label{lem:far}
After the three prescribed passes at scale $s$, and throughout the remaining passes of that scale, $v_{ij}=0$ whenever $j-i\ge8s^{3/2}$.
\end{lemma}
\begin{proof}
Recall that $G=7s/4$ and $m=\lceil\sqrt G\rceil$. Let $B$ be the array after the passes at gaps $G+m$, $G+1$, and $G$. Each pass makes the array sorted at its own gap, and Lemma~\ref{lem:preserve} ensures that the subsequent passes preserve this property. Consequently, $B$ is simultaneously sorted at all three gaps.

Suppose a distance $d$ can be written as
\[
                 d=uG+v(G+m)+w(G+1),\qquad u,v,w\in\mathbb Z_{\ge0}.
\]
For any positions $i<j$ with $j-i=d$, start at $i$ and take $u$ steps of length $G$, then $v$ steps of length $G+m$, then $w$ steps of length $G+1$. At each step, sortedness at the corresponding gap ensures that the key rank cannot decrease. Then we have $B_i\le B_j$. It remains to show that every sufficiently large distance admits such a representation.

Fix an integer $d\ge3G^{3/2}$. Divide first by $G$, then divide the remainder by $m$, to obtain
\[
                 d=qG+r,\quad 0\le r<G,\qquad
                 r=am+b,\quad 0\le b<m,
\]
where $q,a,b$ are nonnegative integers. The choice $m=\lceil\sqrt G\rceil$ gives
\[
                 a\le\frac{r}{m}<\frac{G}{m}\le\sqrt G,
                 \qquad b\le m-1<\sqrt G.
\]
Thus $a+b<2\sqrt G$. On the other hand,
\[
                 q=\left\lfloor\frac dG\right\rfloor
                 \ge3\sqrt G-1\ge2\sqrt G,
\]
where the last inequality uses $G\ge1$. In particular, $q\ge a+b$. Start with $q$ steps of length $G$. Replacing $a$ of them by steps of length $G+m$ and $b$ others by steps of length $G+1$ adds exactly $am+b=r$ to the total distance. Note that, $q\geq a+ b$ makes this replacement feasible.

Finally, $3G^{3/2}=3(7/4)^{3/2}s^{3/2}<8s^{3/2}$. Hence every distance $j-i\ge8s^{3/2}$ has the required representation, so $B_i\le B_j$ and its inversion weight is zero. Every later pass preserves sortedness at the three fixed gaps by Lemma~\ref{lem:preserve}, so the same path argument applies throughout the rest of the scale.
\end{proof}

\subsection{Capturing Intermediate Inversions}\label{sec:routing}
Fix a scale $s$ and any of its eight intervals $\I\triangleq \{a,\ldots,b\}$, of width $w\triangleq b-a+1=s/16$. Thus $s\le a\le b<3s/2$ and $b\le a+w\le 17a/16$. Throughout this section set
\[
                       L\triangleq 2^{17}s,\qquad U\triangleq 8s^{3/2}.
\]
We call a distance $d$ \emph{intermediate} if $L\le d<U$.

We seek ways to split such a distance into two parts. The first part spans $k$ steps of length $h$, and the second spans $k+1$ steps of length $g$, with both gaps in the same candidate interval $\I$. The next lemma guarantees $\Omega(s)$ choices for every intermediate distance.

\begin{lemma}\label{lem:representations}
Fix an integer distance $d$ with $L\le d<U$. Consider representations of $d$ using two candidate gaps $h,g\in\I$ and a positive integer $k$ such that
\[
                              d=kh+(k+1)g.
\]
Let $N_d$ be the number of distinct triples $(k,h,g)$ satisfying these conditions, allowing $h=g$. Then
\[
                         N_d\ge \frac{99}{100}\frac{w^2}{2b}.
\]
\end{lemma}
\begin{proof}
For a fixed integer $k\ge1$, the equation determines $h=\frac{d-g}{k}-g$. The real values of $g$ for which $g,h\in[a,b]$ form the interval
\[
       J_k\triangleq [a,b]\cap\left[\frac{d-kb}{k+1},\,\frac{d-ka}{k+1}\right].
\]
The admissible integers $g$ are exactly those in $J_k$ satisfying $g\equiv d\pmod{k}$. This residue class has spacing $k$, so a nonempty $J_k$ of length $\ell_k$ contains at least $\frac{\ell_k}{k}-1$ admissible values. Each such $g$ uniquely determines $h$ and hence one triple $(k,h,g)$. Put $\ell_k\triangleq 0$ when $J_k$ is empty. Summing the lower bounds over the nonempty intervals gives
\[
\begin{aligned}
 N_d
 &\ge \sum_{\substack{k\ge1\\J_k\ne\varnothing}}
       \left(\frac{\ell_k}{k}-1\right) = \sum_{k\ge1}\frac{\ell_k}{k} -\#\{k\ge1:J_k\ne\varnothing\}.
\end{aligned}
\]
To bound this number, observe that a nonempty $J_k$ requires $(2k+1)a\le d\le(2k+1)b$, since both $h$ and $g$ lie in $[a,b]$. Solving these inequalities for $k$ gives $\frac{d-b}{2b}\le k\le\frac{d-a}{2a}.$
The length of this interval is
\[
             \frac{d-a}{2a}-\frac{d-b}{2b}
             =\frac{d(b-a)}{2ab}=\frac{d(w-1)}{2ab}.
\]
The number of integers in a closed interval is at most its length plus one. Consequently,
\[
               \#\{k\ge1:J_k\ne\varnothing\}
               \le\frac{d(w-1)}{2ab}+1.
\]
Substituting this bound into the preceding sum yields
\begin{equation}\label{eq:count-lengths}
        N_d\ge \sum_{k\ge1}\frac{\ell_k}{k}-\frac{d(w-1)}{2ab}-1 .
\end{equation}

To bound the sum, fix $y\in[a,b]$. By the definition of $J_k$, the condition $y\in J_k$ is equivalent to
\[
                      k\in\left[\frac{d-y}{y+b},\,\frac{d-y}{y+a}\right].
\]
Since $d\ge2^{17}s$ and $b<\frac{3s}{2}$, the lower endpoint $\frac{d-y}{y+b}$ exceeds $1$, so every integer in this interval is a positive index. The interval contains at least its length minus one integers. For each such $k$, the upper endpoint gives $\frac{1}{k}\ge\frac{y+a}{d-y}$. Therefore,
\begin{align*}
   \sum_{k:\,y\in J_k}\frac{1}{k}
      &\ge \left(\frac{d-y}{y+a}-\frac{d-y}{y+b}-1\right)\frac{y+a}{d-y}\\
      &=\frac{w-1}{y+b}-\frac{y+a}{d-y}\\
      &\ge\frac{w-1}{2b}-\frac{2b}{d-b}.
\end{align*}
Integrating over $y\in[a,b]$, an interval of length $b-a=w-1$, gives $\sum_k\frac{\ell_k}{k}\ge \frac{(w-1)^2}{2b}-\frac{2b(w-1)}{d-b}$. Substituting in~\eqref{eq:count-lengths} yields
\[
        N_d\ge\frac{(w-1)^2}{2b}-\frac{2b(w-1)}{d-b}-\frac{d(w-1)}{2ab}-1 .
\]

It remains to check the rounding loss relative to $\frac{w^2}{2b}$. The distance range $2^{17}s\le d<8s^{\frac{3}{2}}$ implies $\sqrt{s}\ge2^{14}$. Using $w=\frac{s}{16}$, $a\ge s$, and $b<\frac{3s}{2}$, we obtain
\begin{align*}
   1-\frac{2bN_d}{w^2}
      &\le \frac{2}{w}+\frac{4b^2}{w(d-b)}
                      +\frac{d}{aw}+\frac{2b}{w^2}\\
      &\le 2^{-23}+\frac{144}{2^{17}-\frac{3}{2}}
                      +\frac{1}{128}+\frac{768}{2^{28}} < \frac{1}{100}.
\end{align*}
Rearranging gives $N_d>\frac{99}{100}\frac{w^2}{2b}$, which implies the stated inequality.
\end{proof}

Define the total intermediate weight by $M_s\triangleq \sum_{i<j:\,L\le j-i<U}v_{ij}$, with $M_s=0$ when the range is empty. A representation $j-i=kh+(k+1)g$ gives the path $i\to p\to j$, where $p\triangleq i+kh=j-(k+1)g$. Its two segments are the position pairs $(i,p)$ and $(p,j)$, spanning $k$ steps of gap $h$ and $k+1$ steps of gap $g$, respectively. The positive-part weights satisfy
\begin{equation}\label{eq:tri}
                  v_{ij}\le v_{ip}+v_{pj}\qquad(i<p<j).
\end{equation}

\begin{lemma}[Expected potential decrease]\label{lem:routing}
Fix any permutation $A$ and choose a gap $h$ uniformly from $\I$. Then the expected decrease in potential after this pass satisfies
\begin{equation}\label{eq:route}
       \E_h\!\left[Q(A)-Q(P_h\circ A)\right]
       \ \ge\ \frac{99}{100}\frac{a}{2b}\,M_s
       \ \ge\ \frac{198}{425}\,M_s .
\end{equation}
\end{lemma}
\begin{proof}
We call a position pair $(x,y)$, $1\le x<y\le n$, together with a gap $q\in\I$ dividing $y-x$, a \emph{labeled edge}, and denote it by $((x,y),q)$. Its weight is $v_{xy}$, and it spans $\frac{y-x}{q}$ steps in a $q$-chain. A position pair with different gap labels gives different labeled edges. For each label $q$, the sum of their weights is exactly $W_q(A)$.

For an intermediate pair $i<j$, every representation $j-i=kh+(k+1)g$ gives two labeled edges $((i,p),h)$ and $((p,j),g)$. Averaging~\eqref{eq:tri} over its $N_{j-i}$ representations yields
\[
 v_{ij}\le\frac{1}{N_{j-i}}
       \sum_{\substack{k\ge1,\ h,g\in\I\\j-i=kh+(k+1)g}}
       \left(v_{i,i+kh}+v_{i+kh,j}\right).
\]
Each occurrence of a labeled edge in this average therefore has coefficient $\frac{1}{N_{j-i}}$.

Fix a labeled edge $((x,y),q)$. We bound its number of occurrences across the paths for all intermediate pairs $i<j$ by considering its two possible roles.
\begin{enumerate}[label=(\roman*)]
\item As the first edge, it forces $x=i$, $y=p$, and $q=h$. Its endpoints and label determine $k=\frac{y-x}{q}$, so the second segment has $k+1$ steps. Each choice of $g\in\I$ determines $j\triangleq y+(k+1)g$ and hence at most one path. There are $w$ candidate gaps $g$, giving at most $w$ occurrences in this role.
\item As the second edge, it forces $x=p$, $y=j$, and $q=g$. Now $k+1=\frac{y-x}{q}$ determines $k$. Each choice of $h\in\I$ determines $i\triangleq x-kh$ and hence at most one path. There are $w$ candidate gaps $h$, giving at most $w$ occurrences in this role.
\end{enumerate}
Thus a fixed labeled edge occurs at most $2w$ times. This bound uses the requirement that the two step counts are exactly $k$ and $k+1$.

By Lemma~\ref{lem:representations}, each occurrence has coefficient at most $\frac{200b}{99w^2}$. Its total coefficient, after summing the averaged inequalities over all intermediate pairs, is therefore at most
\[
                     2w\cdot\frac{200b}{99w^2}=\frac{400b}{99w}.
\]
Grouping the terms by their labeled edges gives
\begin{align*}
 M_s
    &\le\frac{400b}{99w}
         \sum_{q\in\I}\sum_{\substack{x<y\\q\mid(y-x)}}v_{xy}=\frac{400b}{99w}\sum_{q\in\I}W_q(A).
\end{align*}
By Lemma~\ref{lem:cost} and the uniform choice of $h$,
\begin{align*}
 \E_h\!\left[Q(A)-Q(P_h\circ A)\right]
    &=\frac{2}{w}\sum_{h\in\I}hW_h(A)\ge\frac{2a}{w}\sum_{h\in\I}W_h(A)\ge\frac{99}{100}\frac{a}{2b}\,M_s\ge\frac{198}{425}\,M_s.
\end{align*}
The last inequality uses $\frac{a}{b}\ge\frac{16}{17}$. When the intermediate range is empty, $M_s=0$ and the bound follows from the nonnegativity of the potential decrease.
\end{proof}

\subsection{Contraction and Total Comparisons}\label{sec:finish}
We combine the weight captured at intermediate distances with the bound on the total weight of short inversions.
\begin{lemma}\label{lem:drift}
Let $\I$ be any of the eight intervals at scale $s$, let $L\triangleq 2^{17}s$, and suppose $A$ has no inversion at distance at least $U\triangleq 8s^{3/2}$. For $h$ uniform on $\I$,
\begin{equation}\label{eq:drift}
                  \E_h Q(P_h \circ A)\le 0.77\,Q(A)+30nL^2 .
\end{equation}
\end{lemma}
\begin{proof}
Recall that $\delta_i=A_i-i$, where the true rank $A_i$ is the final position of the key currently at $i$. If $\delta_i>0$, its final position lies $\delta_i$ places to the right of its current position. If $\delta_i<0$, its final position lies $-\delta_i$ places to the left. The quantities $\pos{\delta_i}$ and $\pos{-\delta_i}$ record these two displacement amounts, respectively.

For $i<j$, the identity $A_i-A_j=\delta_i-\delta_j-(j-i)$ gives $v_{ij}\le\pos{\delta_i}+\pos{-\delta_j}$. In this upper bound, the first term is assigned to the left endpoint $i$, and the second to the right endpoint $j$. For a fixed left endpoint $i$, at most $L$ positions $j$ satisfy $i<j$ and $j-i<L$, so $\pos{\delta_i}$ is counted at most $L$ times. Similarly, for a fixed right endpoint $j$, the term $\pos{-\delta_j}$ is counted at most $L$ times. Therefore,
\begin{align*}
 \sum_{i<j:\,j-i<L}v_{ij}
    &\le\sum_{i<j:\,j-i<L}\left(\pos{\delta_i}+\pos{-\delta_j}\right)\\
    &\le L\sum_i\pos{\delta_i}+L\sum_j\pos{-\delta_j}\\
    &=L\sum_i|\delta_i|=LF(A)\le L\sqrt{nQ(A)}.
\end{align*}
The last inequality follows from Cauchy--Schwarz, since
\[
 F(A)=\sum_{i=1}^n|\delta_i|\cdot1
      \le\sqrt{\sum_{i=1}^n\delta_i^2}\,
           \sqrt{\sum_{i=1}^n1^2}
      =\sqrt{nQ(A)}.
\]
By Lemma~\ref{lem:cost}, the total inversion weight is $\frac{Q(A)}{2}$. Since there are no inversions at distance at least $U$, removing the weight at distances below $L$ leaves the intermediate weight. Hence
\[
                              M_s\ge\frac{Q(A)}{2}-L\sqrt{nQ(A)}.
\]
When $U\le L$, this gives $Q(A)\le4nL^2$ directly.

Lemma~\ref{lem:routing} gives $\E_h[Q(A)-Q(P_h\circ A)]\ge\frac{198}{425}M_s$. Combining this with the bound on $M_s$ and applying $L\sqrt{nQ(A)}\le\frac{Q(A)}{256}+64nL^2$ yields
\[
   \E_h Q(P_h \circ A)\le \left(1-\frac{127}{256}\cdot\frac{198}{425}\right)Q(A)+\frac{12672}{425}\,nL^2
                 \le 0.77\,Q(A)+30nL^2 .\qedhere
\]
\end{proof}

\begin{lemma}\label{lem:scale}
There is an absolute constant $\beta$ such that for every scale $s$ and every array $A$ entering it, the array $A'$ leaving it satisfies
\[
                 \E[Q(A')\mid A]\le Q(A)/8+\beta ns^2 .
\]
\end{lemma}
\begin{proof}
Let $Q_0$ be the potential after the three fixed passes and $Q_t$ the potential after the $t$-th random pass, in execution order. Lemma~\ref{lem:cost} gives $Q_0\le Q(A)$. By Lemma~\ref{lem:far}, all pairs at distance at least $8s^{3/2}$ remain ordered before each of the eight draws. Conditional on the input and all previous draws, collected in $\mathcal F_{t-1}$, the next gap is uniform on its prescribed interval. Lemma~\ref{lem:drift} therefore gives
\[
    \E[Q_t\mid\mathcal F_{t-1}]\le 0.77\,Q_{t-1}+30\cdot2^{34}ns^2.
\]
Iterating the recurrence gives
\[
    \E[Q_8\mid A]\le0.77^8 Q(A)+\frac{30\cdot2^{34}}{1-0.77}\,ns^2.
\]
Since $0.77^8<\frac{1}{8}$, we may take $\beta\triangleq \frac{30\cdot2^{34}}{1-0.77}$.
\end{proof}

\begin{proof}[Proof of Theorem~\ref{thm:main}]
Fix an input permutation. For $n\ge256$ put $D\triangleq \max\{64,8\beta\}$. We claim that the potential $Q_{\rm in}(s)$ on entry to every scale satisfies
\begin{equation}\label{eq:budget}
                        \E Q_{\rm in}(s)\le Dns^2 .
\end{equation}
Initially $n/8<s_0\le n/4$, and $Q\le n^3\le64ns_0^2$. If~\eqref{eq:budget} holds at $s$, taking expectations in Lemma~\ref{lem:scale} gives
\[
     \E Q_{\rm out}(s)\le (D/8+\beta)ns^2
          \le (D/4)ns^2=Dn(s/2)^2 .
\]
This is the entry condition at the next scale and also applies on exit from the last scale.

Each pass at $s$ begins with $Q\le Q_{\rm in}(s)$ pointwise and has $h\ge s$. By Lemma~\ref{lem:cost} and Jensen's inequality its expected comparison count is at most
\[
        n+\frac{2\sqrt n}{s}\E\sqrt{Q_{\rm in}(s)}
        \le n+\frac{2\sqrt n}{s}\sqrt{Dns^2}=(1+2\sqrt D)n .
\]
There are eleven passes per scale and $O(\log n)$ scales. After scale $64$, $\E Q\le Dn\cdot32^2$, so the final $1$-pass costs at most $(1+64\sqrt D)n$ comparisons in expectation. For $2\le n<256$, direct insertion sort costs at most $n^2<256n$. All constants are independent of the input and of $n$. The last gap is $1$, which guarantees correctness for every draw.
\end{proof}

\begin{proof}[Proof of Corollary~\ref{cor:fixed}]
Draw $A$ uniformly and independently of $H$. Theorem~\ref{thm:main} gives
\[
       \E_H\overline C(H)=\frac1{n!}\sum_{A\in\Sn}\E_H C(H,A)=O(n\log n),
\]
so at least one $H_n$ attains this upper bound. Conversely, a fixed sequence induces a binary comparison tree with a different leaf for every input permutation. The average leaf depth under uniform input is at least $\log_2(n!)=\Omega(n\log n)$, by Kraft's inequality and Jensen's inequality. This lower bound applies to $H_n$.
\end{proof}

\section{Constructing a Fixed Gap Sequence}\label{sec:construct}
This section, we give a efficient randomized algorithm to construct a sequence with $O(n\log n )$ average-case complexity for each $n$ with high probability.

The construction selects the gaps during preprocessing by testing candidates on random permutations. For a gap prefix $J$, let $X_J$ be the array obtained by applying $J$ to a uniform permutation, and define $\Phi(J)\triangleq \E Q(X_J)$. Write $Jh$ for $J$ extended by a gap $h$. The score $\Phi(Jh)=\E Q(P_h\circ X_J)$ measures the mean potential after this extension. The algorithm chooses a gap with minimum estimated score in each candidate interval.

\heading{Construction algorithm.}
The input consists of $n\ge2$ and $0<\delta<1$, and the output is a gap sequence $H$.
\begin{enumerate}
\item If $n<256$, return $H\triangleq (1)$. Otherwise, initialize $J$ to the empty sequence and let $s_0$ be the largest power of two at most $\frac{n}{4}$.
\item For each scale $s=s_0,\frac{s_0}{2},\ldots,64$, set $G\triangleq \frac{7s}{4}$ and $m\triangleq \lceil\sqrt G\rceil$, and append $G+m,G+1,G$ to $J$. Set $w\triangleq \frac{s}{16}$ and use the intervals
\[
       \I_t\triangleq \{s+tw,\ldots,s+(t+1)w-1\},\qquad 0\le t<8.
\]
Visit $\I_7,\ldots,\I_0$ in this order. For each current interval $\I$, perform the following steps.
\begin{enumerate}[label=(\alph*)]
\item Generate a fresh batch of
\[
       m_s\triangleq \left\lceil\left(\frac ns\right)^4\right\rceil
           \left\lceil\log_2\frac{4n}{\delta}\right\rceil
\]
independent test permutations using the sampling procedure below. Apply the current prefix $J$ to each, obtaining $X_J^{(1)},\ldots,X_J^{(m_s)}$.
\item For every $h\in\I$, apply one $h$-pass to a copy of each $X_J^{(r)}$ and compute
\[
       \widehat\Phi(Jh)\triangleq \frac{1}{m_s}\sum_{r=1}^{m_s}
                              Q(P_h\circ X_J^{(r)}).
\]
\item Choose a gap $\hat h\in\I$ minimizing $\widehat\Phi(Jh)$, breaking ties by the smallest gap, and append $\hat h$ to $J$.
\end{enumerate}
\item After the last scale, append $1$ and return $H\triangleq J$.
\end{enumerate}

\heading{Sampling the test permutations.}
For $n\ge256$, the total number of test permutations is $M\triangleq 8\sum_{s=s_0,s_0/2,\ldots,64}m_s$, which is fixed by $n$ and $\delta$. Set $b\triangleq \lceil\log_2\frac{2Mn^2}{\delta}\rceil$. To generate one test permutation, assign an independent uniform $b$-bit integer to each index in $\{1,\ldots,n\}$, then list the indices in increasing order of their assigned integers, breaking ties by index. Use independent random bits for all samples. This implements the sampling with a bounded number of fair random bits.

\begin{proof}[Proof of Theorem~\ref{thm:construct}]
For $n<256$, the comparison bound for the returned sequence $(1)$ follows as in Theorem~\ref{thm:main}. Assume $n\ge256$. The length and decreasing order of the output follow from Section~\ref{sec:schedule}, since every selected gap belongs to its prescribed interval.

The probability that any of the $M$ samples has a tie among its assigned integers is at most
\[
                         M\binom n2 2^{-b}\le\frac{\delta}{2}.
\]
Conditional on no such tie, all test permutations are independent and uniform. We analyze the construction under this conditioning.

\heading{Accuracy needed for one choice.}
The three fixed gaps have already eliminated all inversions at distance at least $8s^{3/2}$. Averaging Lemma~\ref{lem:drift} over $X_J$ gives
\[
   \frac{1}{|\I|}\sum_{h\in\I}\Phi(Jh)\le 0.77\,\Phi(J)+30\cdot2^{34}ns^2 .
\]
Suppose every score estimate differs from its true value by at most $ns^2$. Since the algorithm minimizes $\widehat\Phi(Jh)$, accounting for the estimation errors at its chosen gap $\hat h$ and at a true minimizer gives
\[
     \Phi(J\hat h)\le \min_{h\in\I}\Phi(Jh)+2ns^2
                   \le 0.77\,\Phi(J)+(30\cdot2^{34}+2)ns^2 .
\]
Fix any realization for which all estimates are accurate. Let $J_{\rm in}$ and $J_{\rm out}$ be the prefixes on entry to and exit from scale $s$. The fixed passes can only decrease $\Phi$. Iterating over the eight chosen gaps gives
\[
     \Phi(J_{\rm out})\le 0.77^8\Phi(J_{\rm in})+\beta' ns^2,
     \qquad \beta'\triangleq \frac{30\cdot2^{34}+2}{1-0.77}.
\]
Since $0.77^8<\frac{1}{8}$, taking $D\triangleq \max\{64,8\beta'\}$ in the scale induction and comparison bound of Theorem~\ref{thm:main} yields $\overline C(H)\le c\,n\log n$.

\heading{Accuracy and running time.}
Conditional on the current prefix $J$, the estimate $\widehat\Phi(Jh)$ for a fixed candidate $h$ averages $m_s$ independent observations with mean $\Phi(Jh)$ and range $[0,n^3]$. Hoeffding's inequality~\cite{hoeffding} bounds the probability of error exceeding $ns^2$ by
\[
            2\exp\left(-2m_s\left(\frac sn\right)^4\right)
            \le\frac{\delta}{2n}.
\]
Across the eight intervals at scale $s$ we estimate $8w=\frac{s}{2}$ scores. Summing over the halving scales gives fewer than $s_0\le\frac{n}{4}$ estimates in total. The conditional error bound holds for every realized prefix, so a union bound gives probability at most $\frac{\delta}{2}$ that any estimate is inaccurate, even though earlier batches determine later prefixes. Adding the probability of a tie gives total failure probability at most $\delta$.

Each score can be computed using $m_s$ permutations and at most $O(\log n)$ passes on each. The number of scores, the batch sizes, the total sample count $M$, and the bit length $b$ are polynomial in $n$ and $\log\frac{1}{\delta}$. Thus the algorithm has the claimed polynomial running time and success probability.
\end{proof}


\section*{Declaration of AI use}
The proofs in this paper were obtained by LLM. The authors verified the arguments, worked through the proofs in detail, and polished their exposition and presentation.

{\small
\bibliographystyle{plainurl}
\bibliography{references}
}
\end{document}